\documentstyle[12pt]{article}

\renewcommand{\title}[1]{{\Large\bf\mbox{}\\\medskip#1\bigskip\medskip\\}}
\renewcommand{\author}[1]{{\large #1\smallskip\\}}
\newcommand{\address}[1]{{\em #1\medskip\\}}

\def\be{\begin{equation}}
\def\ee{\end{equation}}

\def\l{{\lambda}}

\begin{document}
\begin{center}

\title{Spanning Trees on Hypercubic Lattices\\
and Non-orientable Surfaces}
\author{W.-J. Tzeng$^1$ and F. Y. Wu$^{2}$}
\address{$^1$Department of Physics, Tamkang University \\
Tamsui, Taipei, Taiwan 251, R. O. C.\\
$^2$Department of Physics, Northeastern University\\ 
Boston, Massachusetts 02115}

\begin{abstract}
We consider the problem of enumerating spanning trees on lattices.
Closed-form expressions are obtained for the spanning tree 
generating function for a hypercubic lattice of size 
$N_1\times N_2\times\cdots\times N_d$ in $d$ dimensions under free,
periodic, and a combination of free and periodic boundary conditions.
Results are also obtained for a simple quartic
net embedded on two non-orientable surfaces,
a M\"obius strip and  the Klein bottle.
Our results are based on the use of a formula 
expressing the spanning tree generating function in terms of the
eigenvalues of an associated tree matrix.  
An elementary derivation of this formula is given.
\end{abstract}
\end{center}


\section{INTRODUCTION}

The problem of enumerating spanning trees on a graph was first 
considered by Kirchhoff \cite{kirk} in his analysis of
electrical  networks.
Consider a graph $G=\{V,E\}$ consisting of 
a vertex set $V$ and an edge set $E$.  
We shall assume that $G$ is connected.
A subset of edges $T\subset E$ is a  spanning tree if 
it has $|V|-1$ edges with at least one edge incident at each vertex.
Therefore $T$ has no cycles. In ensuing  discussions we shall use $T$ 
to also 
denote the spanning tree.

Number the vertices from 1 to $|V|$ and associate 
to the edge $e_{ij}$ connecting vertices $i$ and $j$
a weight $x_{ij}$,
with the convention of $x_{ii}=0$.
The enumeration of spanning trees concerns with the 
evaluation of the tree generating function
\begin{equation}
T(G;\{x_{ij}\}) = \sum_{T\subseteq E}\, \prod _{e_{ij}\in  T} x_{ij}, 
\label{gen}
\end{equation}
where the summation is taken over all spanning trees $T$.
Particularly, the number of spanning trees on $G$ is obtained 
by setting $x_{ij}=1$ as
\begin{equation}
N_{SPT} (G) =  T(G;1). \label{spt}
\end{equation}

Considerations of spanning tree also arise in statistical physics  
\cite{temperley} in the enumeration of 
close-packed dimers (perfect matchings) \cite{temperley1}.  
Using a similar consideration, for example, one of us \cite{wu79} has 
evaluated the number of spanning trees for the simple quartic, 
triangular and honeycomb lattices in the limit of $|V|\to \infty$.
In this {\it Letter} we report new results on the evaluation of
the generating function Eq.\ (\ref{gen}) for {\it finite}  
hypercubic lattices in arbitrary dimensions.
Results are also obtained for a simple quartic net embedded on
two non-orientable surfaces, 
 the  M\"obius strip and the Klein bottle. 
As the main formula used in this Letter is a relation expressing 
the tree generating function in terms of the eigenvalues of 
an associated tree matrix, for completeness 
we give an elementary derivation of this formula.
 
\section{THE TREE MATRIX}
For a given graph $G=\{V,E\}$ consider a $|V|\times |V|$  matrix 
{\bf M}$(G)$ with elements
\begin{equation}
M_{ij}(G)=\left\{\begin{array}{ll}
\sum _{k=1}^{|V|} x_{ik},~~&i=j=1,2,\cdots,|V|\\
-x_{ij},&\mbox{if vertices $i$, $j$, $i\not= j$, are connected 
by an edge}\\
0,&\mbox{otherwise.}
\end{array}\right. 
\label{matrix}
\end{equation}
We shall refer to {\bf M}$(G)$ simply as the tree matrix.
It is well-known \cite{broooks,harary} that the tree generating 
function, Eq.\ (\ref{gen}), is given by the cofactor of any element 
of the tree matrix, and that the cofactor is the same for all elements. 
Namely, we have the identity
\begin{equation}
T(G;\{x_{ij}\}) = \mbox{the cofactor of {\sl any} element 
of the matrix {\bf M}(G)}. 
\label{cofactor}
\end{equation}

The tree generating function can also be expressed 
in terms of the eigenvalues of the tree matrix {\bf M}$(G)$.
We give here an elementary derivation of this result which we use
 in subsequent sections.

Let ${\bf M}(G)$ be the tree matrix of a graph $G=\{V,E\}$. 
Since the sum of all elements in a row of ${\bf M}(G)$ 
equals to zero, ${\bf M}(G)$ has 0 as an eigenvalue and,
by definition, we have 
\be
{\rm det}\left| M_{ij}(G) -\l\delta_{ij}\right| = -\l F(\l) \label{det}
\ee
where
\be
F(\l) = \prod_{i=2}^{|V|} (\l_i-\l), \label{f}
\ee
 $\l_2, \l_3, \ldots, \l_{|V|}$ being the remaining eigenvalues.
 
Now the sum of all elements in a row of the determinant
$\bigl| M_{ij}(G) -\l\delta_{ij}\bigr| $ is $-\l$.  
This permits us to replace the first column of 
${\rm det}\bigl| M_{ij}(G) -\l\delta_{ij}\bigr| $ by a column of  
elements $-\l$ without affecting its value.
Next we carry out a Laplace expansion of the resulting determinant
along the modified  column, obtaining 
\be
{\rm det} \bigl| M_{ij}(G) -\l\delta_{ij}\bigr| 
 = -\l \sum_{i=1}^{|V|} C_{i1}(\l), \label{cof}
\ee
where $C_{i1}(\l)$ is the cofactor of the ($i1$)-th element
of the determinant.
Combining Eqs.\ (\ref{det})-(\ref{cof}), we are led to the identity
\be
F(\l)= \sum_{i=1}^{|V|} C_{i1}(\l). \label{cf}
\ee

Now, $C_{i1}(0)$ is precisely the cofactor of the ($i1$)-th 
element of ${\bf M}(G)$ which, by Eq.\ (\ref{cofactor}), is equal 
to the tree generating function $T(G;\{x_{ij}\})$. 
It follows that, after setting $\l=0$ in Eq.\ (\ref{cf}), 
we obtain an expression giving the tree generating
function in terms of the eigenvalues of the tree matrix [7, p.\ 39] 
\begin{equation}
T(G;\{x_{ij}\})= {1\over {|V|}}\prod_{i=2}^{|V|}\lambda_i.
\label{key}
\end{equation}
This result can also be deduced by considering the tree matrix
of a graph obtained from $G$ by adding an auxiliary vertex connected to
all vertices with edges of weight $x$, followed by
taking the limit of $x\to 0$ \cite{kpw}.


\section{HYPERCUBIC LATTICES}
We now deduce the closed-form expression for the tree 
generating function for a hypercubic lattice in $d$ dimensions 
under various boundary conditions.

\smallskip
\noindent{\it 3.1. Free boundary conditions}

\noindent THEOREM 1.~~
{\it Let ${\bf Z}_d$ be a $d$-dimensional hypercubic lattice of size
$N_1 \times N_2 \times \cdots \times N_d$ with edge weights $x_i$ along 
the $i$th direction, $i=1, 2, \ldots, d$.
The tree generating function for ${\bf Z}_d$ is
\begin{eqnarray}
T({\bf Z}_d;\{x_i\}) &=&
\frac{2^{{\cal N}-1}}{\cal N}\prod_{n_1=0}^{N_1-1}\cdots
\prod_{n_d=0}^{N_d-1}
\left[\sum_{i=1}^dx_i \left(1-\cos\frac{n_i\pi}{N_i}\right)\right],\\
&&~~~~~~~~~~~~~~~~~~~~~~~(n_1,\ldots,n_d)\not=(0,\ldots,0),\nonumber
\end{eqnarray}
where ${\cal N} =N_1N_2\cdots N_d$.
}
\smallskip

\noindent  PROOF.

The tree matrix of ${\bf Z}_d$  assumes the form of a linear combination of 
direct products of smaller matrices, 
\begin{eqnarray}
{\bf M}({\bf Z}_d) &=& \sum_{i=1}^d x_i\big[2I_{N_1} \otimes I_{N_2}
\otimes \cdots \otimes I_{N_d}\nonumber\\
&-&I_{N_1} \otimes \cdots \otimes I_{N_{i-1}}
\otimes H_{N_i} \otimes I_{N_{i+1}} \otimes \cdots \otimes I_{N_d}
\big],
\end{eqnarray}
where $I_N$ is an $N \times N$ identity matrix and $H_N$ is the 
$N \times N$ tri-diagonal matrix
\begin{equation}
H_N=\left(\begin{array}{cccccccc}
1&1&0&0&\cdots&0&0&0\\
1&0&1&0&\cdots&0&0&0\\
0&1&0&1&\cdots&0&0&0\\
\vdots&\vdots&\vdots&\vdots&\ddots&\vdots&\vdots&\vdots\\
0&0&0&0&\cdots&1&0&1\\
0&0&0&0&\cdots&0&1&1
\end{array}
\right).
\end{equation} 
 
It is readily verified that $H_N$ is diagonalized by the 
similarity transformation
\begin{equation}
S_NH_NS_N^{-1} = \Lambda_N,
\label{sh}
\end{equation}
where $S_N$ and $S_N^{-1}$ are $N\times N$ matrices with elements
\begin{eqnarray}
\left(S_N\right)_{mn} &=&\left(S_N^{-1}\right)_{nm} 
=\sqrt{\frac{2}{N}}\cos\left[(2n+1)
\left(\frac{m\pi}{2N}\right)\right]+\left(\sqrt{\frac{1}{N}}
-\sqrt{\frac{2}{N}}\right)\delta_{m,0},\nonumber\\
&&m,n=0, 1,\ldots,N-1,
\end{eqnarray}
and $\Lambda_N$ is an $N \times N$ diagonal matrix with diagonal elements
\begin{equation}
\lambda_n=2
\cos\frac{n\pi}{N},~~~n=0,1,\ldots,N-1.
\end{equation}
Here $\delta_{m,n}$ is the Kronecker delta.
It follows that ${\bf M}({\bf Z}_d)$ is diagonalized by
the similarity transformation
\begin{equation}
{\bf S}_{\cal N}{\bf M}({\bf Z}_d){\bf S}_{\cal N}^{-1}
={\boldmath \Lambda}_{\cal N},
\label{sm}
\end{equation}
where 
\begin{equation}
{\bf S}_{\cal N} = S_{N_1} \otimes S_{N_2} \otimes \cdots \otimes S_{N_d},
\end{equation} 
and ${\boldmath \Lambda}_{\cal N}$ is
an ${\cal N}\times{\cal N}$ diagonal matrix with diagonal elements
\begin{equation}
\lambda_{n_1,\ldots,n_d} = 2\sum_{i=1}^dx_i
\left[1-\cos\frac{n_i\pi}{N_i}\right],
~~~n_i=0,1,\ldots,N_i-1.  \label{freeeigen}
\end{equation}

Now, we have $\lambda_{n_1,\ldots,n_d} =0$ for 
$n_1 =n_2 =\cdots =n_d =0$. 
This establishes Theorem\ 1 after using  Eq.\ (\ref{key}).
\hspace*{\fill} Q.E.D.
\smallskip

\noindent REMARK.~~ 
{\it The result Eq.\ (\ref{freeeigen})
generalizes the   $d=2$ eigenvalues of {\bf M}$({\bf Z}_2)$    
for $x_i=1$ reported in} [7, p.\ 74].
\smallskip

\noindent{\it 3.2. Periodic boundary conditions}

In applications in physics one often requires periodic boundary 
conditions depicted by the  condition
 that two ``boundary'' vertices at coordinates
$(\ldots, n_i=1, \ldots)$ and $(\ldots, n_i=N_i, \ldots)$, 
$i=1,2,\ldots,d$, are connected by an extra edge.
This leads to a lattice ${\bf Z}_d^{\rm Per}$ which is a regular graph
with degree $2d$ at all vertices.
For $d=2$, for example, ${\bf Z}_2^{\rm Per}$ can be regarded as 
being embedded on the surface of a torus.
\smallskip

\noindent THEOREM 2.~~
{\it Let ${\bf Z}_d^{\rm Per}$ be a hypercubic lattice in $d$ dimensions
of size $N_1\times N_2 \times \cdots \times N_d$ with edge weights $x_i$ 
along the $i$th direction, $i=1,2,...,d$
with periodic boundary conditions. The
tree generating function for ${\bf Z}_d^{\rm Per}$ is
\begin{eqnarray}
T({\bf Z}_d^{\rm Per};\{x_i\}) &=&\frac{2^{{\cal N}-1}}{\cal N}
\prod_{n_1=0}^{N_1-1}\cdots\prod_{n_d=0}^{N_d-1}
\left[\sum_{i=1}^dx_i \left(1-\cos\frac{2n_i\pi}{N_i}\right)\right],
\label{treeper}\\
&&~~~~~~~~~~~~~~~~~~~~~~~~(n_1,\ldots,n_d)\not=(0,\ldots,0).\nonumber
\end{eqnarray}
}
\smallskip

\noindent PROOF.

The tree matrix assumes the form 
\begin{eqnarray}
{\bf M}({\bf Z}_d^{\rm Per})&=&\sum_{i=1}^d x_i\big[2I_{N_1} \otimes 
I_{N_2} \otimes \cdots \otimes I_{N_d}\nonumber\\
&-&I_{N_1} \otimes \cdots \otimes I_{N_{i-1}} \otimes G_{N_i} 
\otimes I_{N_{i+1}} \otimes \cdots \otimes I_{N_d}\big],
\end{eqnarray}
where $G_N$ is the $N \times N$ cyclic matrix
\begin{equation}
G_N=\left(\begin{array}{cccccccc}
0&1&0&0&\cdots&0&0&1\\
1&0&1&0&\cdots&0&0&0\\
0&1&0&1&\cdots&0&0&0\\
\vdots&\vdots&\vdots&\vdots&\ddots&\vdots&\vdots&\vdots\\
0&0&0&0&\cdots&1&0&1\\
1&0&0&0&\cdots&0&1&0
\end{array}
\right).
\end{equation} 

As in Eq.\ (\ref{sm}), the matrix ${\bf M}({\bf Z}_d^{\rm Per})$ can be 
diagonalized by a similarity transformation generated by 
\begin{equation}
{\bf R}_{\cal N} = R_{N_1} \otimes R_{N_2} \otimes \cdots \otimes R_{N_d}, 
\end{equation}
where $R_N$ is an $N\times N$ matrix with elements
\begin{equation}
\left(R_N\right)_{nj} =\left(R_N^{-1}\right)^*_{jn} 
=N^{-1/2}e^{i2\pi jn/N},
\end{equation}
where $^*$ denotes the complex conjugate,
yielding eigenvalues of $G_N$ as
\begin{equation}
\lambda_n=2\cos\frac{2n\pi}{N}, 
~~~n=0,1,\ldots,N-1.  \label{eigenper}
\end{equation}
This establishes Theorem\ 2 after using
Eq.\ (\ref{key}).
\hspace*{\fill} Q.E.D.
\smallskip

\noindent{\it 3.3. Periodic boundary conditions along $m \le d$ directions}

\noindent COROLLARY.~~
{\it 
Let
${\bf Z}_d^{{\rm Per}(m)}$ be a hypercubic lattice in $d$ dimensions
of size $N_1\times N_2\times \cdots \times N_d$ with
periodic boundary conditions  in 
directions $1,2,\ldots,m\le d$ and free boundaries
in the remaining $d-m$ directions. The tree generating function
is
\begin{eqnarray}
&&T({\bf Z}_d^{{\rm Per}(m)};\{x_i\})=
\frac{2^{{\cal N}-1}}{\cal N}\prod_{n_1=0}^{N_1-1}\cdots\prod_{n_d=0}^{N_d-1}
\left[\sum_{i=1}^mx_i
\left(1-\cos\frac{2n_i\pi}{N_i}\right)
\right.\nonumber\\
&&+\left.\sum_{i=m+1}^dx_i
\left(1-\cos\frac{n_i\pi}{N_i}\right)\right],
~~~~~~~~(n_1,\ldots,n_d)\not=(0,\ldots,0).
\end{eqnarray}
}

\section{THE M\"OBIUS STRIP AND THE KLEIN BOTTLE}
Due to the interplay with the conformal field theory 
\cite{bcn}, it is of current interest in statistical physics to study
lattice systems on non-orientable surfaces \cite{luwu,bs}.
Here, we consider two such surfaces,
the M\"obius strip and the Klein bottle, and obtain the respective
tree generating functions.

\smallskip
\noindent{\it 4.1. The M\"obius strip}

\noindent  THEOREM 3.~~
{\it Let ${\bf Z}_2^{\rm Mob}$ be an
 $M \times N$ simple quartic net embedded on a M\"obius
strip forming a M\"obius net  of width $M$
and  twisted in the direction $N$, with 
edge weights $x_1$ and $x_2$ 
along directions $M$ and $N$ respectively.
The tree generating function for
${\bf Z}_2^{\rm Mob}$ is
\begin{eqnarray}
&&T({\bf Z}_2^{\rm Mob};\{x_1,x_2\})=\frac{2^{MN-1}}{MN}
\prod_{m=0}^{M-1}\prod_{n=0}^{N-1}\left[x_1\left(1
-\cos\frac{m\pi}{M}\right)\right.\nonumber\\
&&~~~~~~~~-\left.x_2\left(1-\cos\frac{4n-3-(-1)^m}{2N}\pi\right)
\right],~~~(m,n)\not=(0,0).
\label{tmob}
\end{eqnarray}
}
\smallskip

\noindent PROOF.

Specifically, let   the the two vertices at coordinates 
$\{m,1\}$ and $\{M-m,N\},m=1,2,\cdots,M$ be connected with a 
lattice edge of weight $x_2$.
Then the tree matrix assumes the form
\begin{equation}
{\bf M}({\bf Z}_2^{\rm Mob})=2(x_1+x_2)I_M\otimes I_N
-x_1H_M\otimes I_N -x_2\big[I_M\otimes F_N +J_M\otimes K_N\big],
\end{equation}
where
\[
F_N=\left(\begin{array}{cccccccc}
0&1&0&0&\cdots&0&0&0\\
1&0&1&0&\cdots&0&0&0\\
0&1&0&1&\cdots&0&0&0\\
\vdots&\vdots&\vdots&\vdots&\ddots&\vdots&\vdots&\vdots\\
0&0&0&0&\cdots&1&0&1\\
0&0&0&0&\cdots&0&1&0
\end{array}
\right),
~~J_M=\left(\begin{array}{cccccc}
0&0&\cdots&0&0&1\\
0&0&\cdots&0&1&0\\
0&0&\cdots&1&0&0\\
\vdots&\vdots&\ddots&\vdots&\vdots&\vdots\\
0&1&\cdots&0&0&0\\
1&0&\cdots&0&0&0
\end{array}
\right),
\] 
\[
K_N=\left(\begin{array}{cccccc}
0&0&0&\cdots&0&1\\
0&0&0&\cdots&0&0\\
0&0&0&\cdots&0&0\\
\vdots&\vdots&\vdots&\ddots&\vdots&\vdots\\
0&0&0&\cdots&0&0\\
1&0&0&\cdots&0&0
\end{array}
\right).
\] 
Since $H_M$ and $J_M$ commute, they can be 
simultaneously diagonalized by
applying the similarity transformation Eq.\ (\ref{sh}).
The transformed matrix 
${\bf S}_{\cal N}{\bf M}({\bf Z}_2^{\rm Mob}){\bf S}_{\cal N}^{-1}$
is ``block diagonal'' with $N\times N$ blocks
\begin{equation}
2\left(x_1-x_1\cos\frac{m\pi}{M}+x_2\right)I_N
-x_2\left(F_N+(-1)^{m}K_N\right),
~~m=0,1,\ldots,M-1.
\label{dmob}
\end{equation}
Now, the eigenvalues of $G_N=F_N+K_N$ and $F_N-K_N$ are,
respectively, $2\cos[2(n+1)\pi/N]$ and
$2\cos[(2n+1)\pi/N], n=0,1,...,N-1$.
Theorem\ 3 is established by combining these results with
Eq.\ (\ref{key}).
\hspace*{\fill} Q.E.D.
\smallskip

\noindent  REMARK.~~
{\it For $M=2$ and $x_1=x_2=1$, Eq.\ (\ref{tmob})
gives the number of spanning trees on a $2\times N$ 
M\"obius ladder as 
\begin{eqnarray}
N_{SPT}&=&\frac{1}{2N}\prod_{j=1}^{2N-1}
\left[3-(-1)^j-2\cos\frac{j\pi}{N}\right] \nonumber\\
& =&\frac{N}{2}\left[  2+ (2+\sqrt{3})^N + (2-\sqrt{3})^N \right] .
\end{eqnarray}
These two equivalent expressions
have previously been given by [7, p.218] and 
by Guy and Harary \cite{gh}, respectively.
}

\smallskip
\noindent{\it 4.2. The Klein bottle}

The embedding of an $M \times N$ simple quartic net on a Klein
bottle is accomplished by further imposing a periodic boundary
condition to ${\bf Z}_2^{\rm Mob}$ in the M direction, namely,
by connecting vertices of ${\bf Z}_2^{\rm Mob}$ at coordinates
$\{1,n\}$ and $\{M,n\}$, $n=1,2,\ldots,N$ with an edge of 
weight $x_1$.
This leads to a lattice ${\bf Z}_2^{\rm Klein}$ of the topology
of a Klein bottle.
\smallskip

\noindent THEOREM 4~~
{\it The tree generating function for ${\bf Z}_2^{\rm Klein}$ 
(described in the above) is
\begin{eqnarray}
&&T({\bf Z}_2^{\rm Klein};\{x_1,x_2\})=\frac{2^{MN-1}}{MN}
\left[\prod_{n=1}^{N-1}2x_2\left(1-\cos\frac{2n\pi}{N}\right)\right]
\nonumber\\
&&~~~~\times\prod_{m=1}^{\left[\frac{M-1}{2}\right]}\prod_{n=0}^{2N-1}
\left[2x_1\left(1-\cos\frac{2m\pi}{M}\right)+
2x_2\left(1-\cos\frac{n\pi}{N}\right)\right]\nonumber\\
&&~~~~\times\left\{\begin{array}{ll}
\prod_{n=0}^{N-1}\left[4z_1-2z_2\left(1-\cos\frac{(2n+1)\pi}{N}\right)
\right],~~~&\mbox{\rm for $M$ even}\\
1,&\mbox{\rm for $M$ odd,}
\end{array}
\right.
\label{tklein}
\end{eqnarray}
where $[n]$ is the integral part of $n$.
}
\smallskip

\noindent PROOF.

The tree matrix of ${\bf Z}_2^{\rm Klein}$ assumes the form
\begin{equation}
{\bf M}({\bf Z}_2^{\rm Klein})=2(x_1+x_2)I_M\otimes I_N
-x_1G_M\otimes I_N -x_2\big[I_M\otimes F_N +J_M\otimes K_N\big].
\label{mklein}
\end{equation}
To obtain its eigenvalues, we
 first apply the similarity transformation generated by $R_M$
in the $M$ subspace. 
While this diagonalizes $G_M$ with eigenvalues 
$2\cos(2m\pi/M)$, $m=0,1,\ldots,M-1$, it transforms the matrix $J_M$ into
\begin{equation}
R_MJ_MR_M^{-1}=\left(\begin{array}{ccccccc}
1&0&0&\cdots&0&0&0\\
0&0&0&\cdots&0&0&\omega\\
0&0&0&\cdots&0&\omega^2&0\\
\vdots&\vdots&\vdots&\ddots&\vdots&\vdots&\vdots\\
0&0&\omega^{M-2}&\cdots&0&0&0\\
0&\omega^{M-1}&0&\cdots&0&0&0
\end{array}
\right),
\end{equation} 
where $\omega =e^{i2\pi/M}$, and thus ${\bf M}({\bf Z}_2^{\rm Klein})$ into
\begin{equation}
\left(\begin{array}{ccccccc}
A_0+B_0&0&0&\cdots&0&0&0\\
0&A_1&0&\cdots&0&0&B_1\\
0&0&A_2&\cdots&0&B_2&0\\
\vdots&\vdots&\vdots&\ddots&\vdots&\vdots\\
0&0&B_{M-2}&\cdots&0&A_{M-2}&0\\
0&B_{M-1}&0&\cdots&0&0&A_{M-1}
\end{array}
\right),
\label{dklein}
\end{equation} 
where $A_m$ and $B_m$ are $N\times N$ matrices
given by
\begin{eqnarray}
A_m&=&2\left[x_1+x_2-x_1\cos\frac{2m\pi}{M}\right]
I_N-x_2F_N,\nonumber \\
B_m&=&-\omega^mx_2K_N,\hskip 1cm m=0,1,\ldots,M-1.
\end{eqnarray}
The matrix Eq.\ (\ref{dklein}) is block diagonal
with blocks
\begin{eqnarray}
 U_N(0)&=&A_0+B_0=x_2(2I_N-G_N)\nonumber \\ 
U_{2N}(m) &=&\left(\begin{array}{cc}A_m&B_m\\B_{M-m}&A_{M-m}
\end{array}\right), \ \ m=1,2,\cdots, \left[{{M-1}\over 2}\right] 
\end{eqnarray}
and for $M=$ even,
\begin{eqnarray}
U_N(M/2)& =& A_{M/2} +B_{M/2},\nonumber\\ 
&=& 2(2x_1 + x_2)I_N - x_2(F_N-K_N),~~ 
\label{u}
\end{eqnarray}
where the subscripts of the $U$ matrices denote the matrix dimensions.
It follows that  we need only to find the eigenvalues of the $U$ matrices.

Eigenvalues of $U_N(0)$ and $U_N(M/2)$ can be deduced from
those of $G_N$ and $F_N-K_N$.  Furthermore,   eigenvalues of $U_{2N}(m)$
are  obtained from those of $G_N$ after applying the similarity transformation
\begin{equation}
T_{2N}(m) U_{2N}(m) T^{-1}_{2N}(m) = 2\left(
x_1 + x_2 - x_1 \cos {{2m\pi}\over m} \right) I_{2N} - x_2 G_{2N}
\end{equation}
where
\begin{equation}
T_{2N}(m)=
\left(\begin{array}{cc}I_N&0\\  0&\omega^{-m}I_N 
\end{array}\right).
\end{equation}
Combining these results  with
Eq.\ (\ref{key}), we are led to Eq.\ (\ref{tklein}) and the theorem.
\hspace*{\fill} Q.E.D.

\section*{Acknowledgment}

We thank T. K. Lee for the hospitality at the Center for 
Theoretical Sciences where this research is carried out.
We are grateful to L. H. Kauffman for a useful conversation and to 
R. Shrock for calling our attention to Refs.
\cite{spectra} and \cite{gh}.
The work of FYW is supported in part by NSF Grant DMR-9614170, and
the support of the National Science Council, Taiwan, is also gratefully
acknowledged.

\end{document}